\documentclass{jpsj2}

\title{%
$^{77}$Se NMR evidence for the Jaccarino-Peter mechanism in the
field induced superconductor, $\lambda$(BETS)$_2$FeCl$_4$ }

\author{%
Ko-ichi \textsc{Hiraki}$^{1,7}$\thanks{E-mail address:
ko-ichi.hiraki@gakushuin.ac.jp}, Hadrien \textsc{Mayaffre}$^{1}$,
Mladen \textsc{Horvati\'{c}}$^{2}$, Claude
\textsc{Berthier}$^{1,2}$, Shinya \textsc{Uji}$^{3}$, Takahide
\textsc{Yamaguchi}$^{3}$, Hisashi \textsc{Tanaka}$^{4}$, Akiko
\textsc{Kobayashi}$^{5}$, Hayao
\textsc{Kobayashi}$^{6}$\thanks{Present address: Department of
Humanities and Sciences, Nihon University, Tokyo} and Toshihiro
\textsc{Takahashi}$^{7}$ }

\inst{%
$^{1}$Laboratoire de Spectrom\'{e}trie Physique, BP87, 38402 St. Martin d'H\`{e}res, France \\
$^{2}$Grenoble High Magnetic Field Laboratory, BP166, 38042 Grenoble, France \\
$^{3}$National Research Institute for Metals, Tsukuba, Ibaraki, 305-0003 \\
$^{4}$National Institute of Advanced Industrial Science and Technology (AIST), Ibaraki 305-8561 \\
$^{5}$Department of Humanities and Sciences, Nihon University, Tokyo 156-8550 \\
$^{6}$Institute for Molecular Sciences, Aichi 444-8585 \\
$^{7}$Department of Physics, Gakushuin University, Tokyo 171-8588
}

\recdate{\today}

\abst{%
 We have performed $^{77}$Se NMR on a single crystal sample
of the field induced superconductor $\lambda$-(BETS)$_{2}$FeCl$_{4}$. Our
results obtained in the paramagnetic state provide a microscopic insight on the
exchange interaction $J$ between the spins \textbf{s} of the BETS $\pi$
conduction electrons and the Fe localized $d$ spins \textbf{S}. The absolute
value of the Knight shift \textbf{K}  decreases  when the polarization of the
Fe spins increases. This reflects the ``negative'' spin polarization of the
$\pi$ electrons through the exchange interaction $J$. The value of $J$ has been
estimated from the temperature and  the magnetic field dependence of \textbf{K}
and found in good agreement with that deduced from transport measurements (L.
Balicas \textit{et al}. Phys. Rev. Lett. \textbf{87}, 067002 (2001)). This
provides a direct microscopic evidence that the field induced superconductivity
is due to the compensation effect predicted by Jaccarino and Peter (Phys. Rev.
Lett. \textbf{9}, 290 (1962)). Furthermore, an anomalous broadening of the  NMR
line has been observed at low temperature, which suggests the existence of
charge disproportionation in the metallic state neighboring the superconducting
phase. }

\kword{ Field Induced Superconductivity, Exchange interaction, NMR }

\begin{document}
\maketitle

\section{Introduction}

\begin{figure}
\begin{center}
\includegraphics[width=7cm]{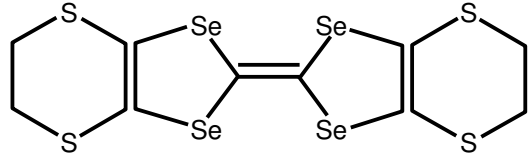}
\end{center}
\caption{ BETS molecule. One of two Se (at random) in each inner five-membered ring is enriched by NMR active $^{77}$Se to nearly 100~\%.

} \label{betsmol}
\end{figure}

Charge transfer complexes based on organic molecules have attracted a huge
amount of interest in the last twenty years due to their low dimensionality and
the possibility to control their electronic properties by modification of base
molecules or pressure. \cite{special_issue} In the case of 2D complexes, a lot
of attention has been paid to superconductivity and its interplay with
electron-electron correlations. In particular, a large number of ET (where ET
stands for the C$_{10}$S$_{8}$H$_{8}$ (bisethylenedithiotetrathiafulvalene)
molecule based superconductors have been extensively studied.\cite{Ishiguro}
Later on, having in mind magnetic properties of these charge transfer
complexes, new materials have been synthesized offering the possibility of an
interaction of the conduction electrons of the $\pi$  band with $d$ localized
electrons.\cite{general_reference} One of these strongly $\pi$-$d$ interacting
system is  $\lambda$-(BETS)$_{2}$FeCl$_{4}$, which is a charge transfer complex
composed of the organic BETS (C$_{10}$S$_{4}$Se$_{4}$H$_{8}$,
bisethylenedithiotetraselenafulvalene, see Fig.~\ref{betsmol}) donor molecule
and magnetic FeCl$_{4}$ (Fe$^{3+}$, $S=5/2$) counter
ion.\cite{kobayashia1993,kobayashih1996} The two-dimensional conducting sheets
are parallel to the crystallographic $ac$ plane and consist of a $\lambda$ type
arrangement of BETS molecules. These sheets are sandwiched by the insulating
FeCl$_{4}$ layers. It turned out that, in addition to its magnetic properties,
the most fascinating property of this compound was again due to its
superconducting properties which appear only in the presence of a strong
applied magnetic field.\cite{uji2001,balicas2001}

In zero external magnetic field ($H_0$)  the system behaves as a metal below
90~K and undergoes a metal-insulator transition around 10~K, accompanied with
an antiferromagnetic ordering.\cite{kobayashia1993,tokumoto1997} The metal
insulator transition temperature decreases with increasing field and above 11~T
the system behaves metallic down to lowest temperature. ESR, magnetization
measurements and theoretical studies reported that the coupling between
delocalized $\pi$ electron having $s$=1/2 spin and the high spin state of Fe
3$d$ spin ($S$=5/2) plays a crucial role in the stabilization of the
antiferromagnetic ordering.\cite{tokumoto1997,brossard1998,hotta2000}

The most remarkable property of $\lambda$-(BETS)$_{2}$FeCl$_{4}$ is the
existence of Field Induced Superconductivity (FISC).\cite{uji2001,balicas2001}
When the magnetic field is applied parallel to the $ac$ plane, the system
becomes superconductor for $H_0 \ge$ 18~T. On increasing $H_0$ the transition
temperature $T_{c}$ grows up to its maximal value $T_c^{max}=4.2$~K at 33~T,
and then decreases and falls down to zero at $H_0=45$~T. To explain this FISC,
the ``compensation'' mechanism predicted by Jaccarino and
Peter\cite{jaccarino1962} (JP) has been proposed. The strong applied magnetic
field polarizes the Fe spins, and the polarized Fe moments
$g\mu_{\rm{B}}$\textbf{S} produce an extra magnetic field on the conduction
electron spins \textbf{s} through the exchange coupling $J$. For
antiferromagnetic $J$ this exchange field is antiparallel (i.e., opposed) to
the applied field, so that the total effective magnetic field can be put to
zero for a certain value of the applied (external) magnetic field. In
$\lambda$-(BETS)$_{2}$FeCl$_{4}$ such complete compensation seems to happen at
33~T, where $T_c^{max}$ is reached. This interpretation in terms of the ``JP
effect'' is reinforced by the observation that the isostructural GaCl$_{4}$
salt, in which the anion is non-magnetic so that there can be no exchange
field, undergoes a superconducting transition under \textit{zero} field at
nearly the same $T_{c}^{\rm{max}}$ value.\cite{kobayashi1997} The JP mechanism
has also been experimentally supported by the transport measurements of Uji
\textit{et al.}, who carried out systematic transport studies on the alloy
system,
$\lambda$-(BETS)$_{2}$Ga$_{1-x}$Fe$_{x}$Cl$_{4}$.\cite{ujiprb2002,ujijpsj2003}
From analysis of the Shubnikov-de Haas oscillations, they have estimated the
exchange field of the $x=1$ (FeCl$_{4}$) salt to 32~T and found it decreasing
with $x$.

From bulk measurements it is difficult to get direct information on the
magnetic behavior of the $\pi$ conduction electrons, since the contribution of
Fe $S=5/2$ spins to the bulk susceptibility is much larger. Only local probes,
like NMR, can provide information on the spin polarization of the $\pi$
conduction electron band. The first reported NMR results on this material have
been obtained on $^{1}$H nuclei.\cite{endo2002,wu2006} Since the coupling of
$^{1}$H nuclei to $\pi$ conduction band is small, while the direct dipolar
coupling to the Fe moments is large, details on the electronic state of the
system could not be obtained. As compared to protons, the Se sites have larger
coupling constant and relatively smaller gyromagnetic ratio. Therefore,
$^{77}$Se NMR measurements at high magnetic field are apparently the best way
to clarify the role of $\pi$ conduction electrons spins in the system. Up to
now, two NMR studies have been performed on the temperature dependence of the
$^{77}$Se shift, and interpreted as a proof of the Jaccarino Peter mechanism
\cite{Hiraki_2006, Wu_2007}. However, a direct measurement of the effective
field experienced by the conduction electrons when the magnetic field is
increased is still missing in \mbox{$\lambda$-(BETS)$_{2}$FeCl$_{4}$}. One
should notice that such a Se NMR evidence for the JP mechanism has been found
in the parent compound \mbox{$\kappa$-(BETS)$_{2}$FeBr$_{4}$} by Fujiyama
\emph{et al.}\cite{fujiyama2006} However, this system is different, since
superconductivity is also present in zero field.\cite{konoike2004}

Here we report $^{77}$Se NMR measurements on a single crystal of
$\lambda$-(BETS)$_{2}$FeCl$_{4}$, which allowed us to detect the spin
polarization of the $\pi$ electrons through the hyperfine coupling to $^{77}$Se
nuclei in the field range  13 - 28 T. We found that the spin polarization
decreases as $H_0$ increases, giving a microscopic evidence that the
Jaccarino-Peter compensation mechanism occurs in this compound. We also
evaluate the coupling constant $J$ between $\pi$ and $d$ spins.

The paper is organized as follows. Experimental details, and characterization
of the sample are described in section~\ref{ex}. In section~\ref{nmr}, we
discuss the NMR shift in presence of an exchange interaction between the
localized spins \textbf{S} and the $\pi$ conduction electrons spins \textbf{s}.
The results and discussion are given in section~\ref{rd}. In subsections
\ref{j} and \ref{k} we discuss the $\pi$-$d$ interaction in the
$\lambda$-(BETS)$_{2}$FeCl$_{4}$ from field and temperature dependence of the
NMR shift. The anomalous line broadening observed at low temperature is
discussed in section~\ref{cd}, where we point out the possibility of charge
disproportionation (CD) in the BETS sheets.

\section{\label{ex}Experimental Details}

Experiments were performed on a $\sim$3 $\times$ 0.05 $\times$ $\sim$0.01
mm$^{3}$ single crystal,
enriched with $^{77}$Se isotope to $\simeq$~50~\% (see caption to Fig.~\ref{betsmol}).
The group symmetry of $\lambda$-(BETS)$_{2}$FeCl$_{4}$ is P$\overline{1}$ (triclinic).
The detailed synthesis procedure is described elsewhere.\cite{takimiya2002,takimiya2003}
Since the natural abundance of $^{77}$Se isotope is only $\sim$7$~\%$, in the
absence of isotopic enrichment the number of observable nuclei would be
$\sim$10$^{15}$, which is a very small number. The
enrichment was thus
a key ingredient to improve the signal to noise (S/N) ratio to a level
compatible with the time-limited experiments in the high-field resistive
magnets. Another key ingredient was to optimize the filling factor of the NMR
coil by making a microcoil wound using 13~$\mu$m insulated Cu wire to obtain
the inner coil diameter of only 75~$\mu$m.

Our crystal was a needle along the $c$ axis, the largest face being the $ac$
plane (\textit{i.e.}, perpendicular to $b^{*}$). The sample and the NMR coil
were mounted on a goniometer with the rotation axis along $c$, allowing the
possibility to vary the direction of $H_{0}$ within the $a^{*}b^{*}$ plane and
to align it along the axis $a'$ which is the intersection of this $a^{*}b^{*}$
plane and the superconducting $ac$ plane. We tried to obtain the initial
orientation such that $H_{0}\parallel$ $a'$ axis, however, the sample being
very small, a precise alignment was very difficult.

\begin{figure}
\begin{center}
\includegraphics[width=11cm]{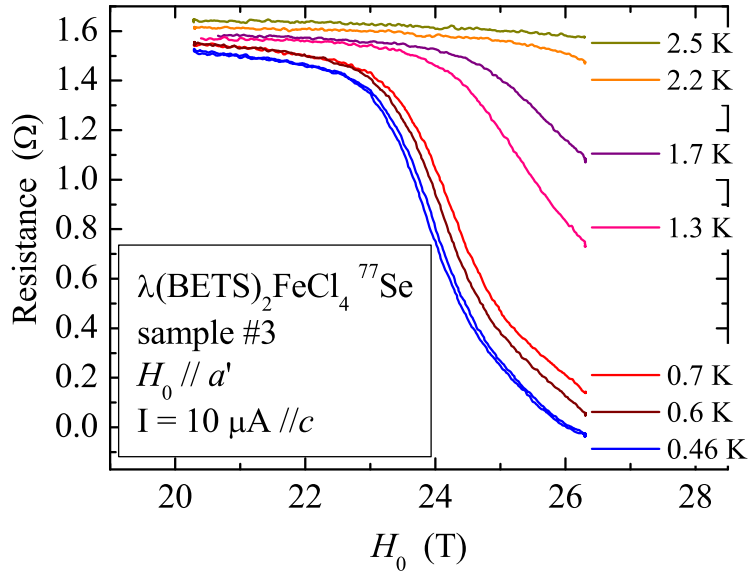}
\end{center}
\caption{ Resistivity as a function of the magnetic field at various
temperature. The sample and that used for NMR measurement sample are part of
the same batch.} \label{transport}
\end{figure}

The field dependence of NMR spectra were measured in the field range between
13~T and 28~T. The NMR spectra were recorded in a superconducting magnet up to
17~T, while in the field range 16-28 T measurements were performed (at 1.5~K)
in a 20 MW resistive magnet of Grenoble High Magnetic Field Laboratory. Spectra
were obtained by the Fourier Transform (FFT) of the spin echo signal at fixed
magnetic field. The linewidths at low temperature were broader than the typical
\textit{rf} (radio frequency) excitation width (of $\sim$0.4~MHz for the pulse
of $\sim$1.2 $\mu$s). In this case the ``frequency sweep'' spectra were
constructed by summing several individual spectra taken at regular frequency
step intervals.\cite{clark1995} The value of the applied magnetic field $H_{0}$
was calibrated using the $^{63}$Cu NMR signal of the NMR coil and the bare
Larmor frequency of $^{77}$Se was taken to be $f_{0} = \gamma H_{0}$, with
$\gamma = 8.127296$~MHz/T, which is equivalent to take  neutral TMTSF as the
reference. \cite{takagi2003} All experiments were carried out in the metallic
state. While we have not detected any superconducting transition while rotating
the sample at 1.5 K, superconducting phase has been observed on samples of the
same batch by transport measurements in the High Field Laboratory at Tsukuba.
However, one can see on Fig.~\ref{transport} that the decrease of the
resistivity at 1.5 K is weak, so that the effect on the NMR spectra is expected
to be negligible.

\section{\label{nmr}NMR background}

The purpose of this work is to determine the polarization of the $\pi$-band
conduction electrons and to relate it to localized spins polarization.  Before
describing the experimental results, we shall first discuss the origin of the
local field at the Se nuclei, and how we can relate it to the quantities of
interest. The minimum starting Hamiltonian describing the interactions between
the nuclear spins $I^{i}$, the conduction electron spins \textbf{s}$^{k}$ and
the localized spins \textbf{S}$^{j}$ at the Fe sites $j$ can be written as
\begin{equation}
\mathcal{H} = \mathcal{H}_{\rm{Z}} + \mathcal{H}_{Is} + \mathcal{H}_{IS} +
\mathcal{H}_{\rm{exch}}~. \label{Ham}
\end{equation}
The first term is the Zeeman interaction for the three types of spins
\begin{equation}
\mathcal{H}_{\rm{Z}} = \sum_{i} - \gamma \hbar I_{z}^{i}(1+ K_{c}^{i}) H_{0} +
\sum_{j} g_{\rm{Fe}}\mu_{\rm{B}}S_{z}^{j}H_{0} + \sum_{k} g_{\pi}\mu_{\rm{B}}
s_{z}^{k} H_{0}~, \label{Zeeman}
\end{equation}
in which $K_{c}^{i}$ is the chemical shift. The last term of Eq.~\ref{Ham}
corresponds to the exchange interaction between the localized and the itinerant
spins
\begin{equation}
\mathcal{H}_{\rm{exch}} = \sum_{j,k} g_{\pi} g_{\rm{Fe}} \mu_{\rm{B}}^{2}
J_{k,j} \textbf{s}^{k}\textbf{S}^{j}~.
\end{equation}
We shall assume that its effect can be expressed as a uniform exchange field
$H_{\rm{exch}}= g_{\rm{Fe}} \mu_{\rm{B}} J \langle S_{z} \rangle$ acting on the
$\pi$ band conduction electrons, that is $\mathcal{H}_{\rm{exch}} \approx
\sum_{k} g_{\pi}\mu_{\rm{B}} s_{z}^{k} H_{\rm{exch}}$. This term is then of the
same form as the Zeeman term for $\pi$ electrons (the last term of
Eq.~\ref{Zeeman}), so that their magnetization $-g_{\pi}\mu_{\rm{B}}\langle
s_{z}\rangle = \chi_{\pi} H_{\rm{eff}}$ is determined by their susceptibility
$\chi_{\pi}$ and the \textit{total} effective magnetic field $H_{\rm{eff}} =
H_{0} + H_{\rm{exch}}$. We note that the Zeeman interaction polarizes Fe spins
antiparallel (negative) to $H_{0}$ (i.e., for positive $H_{0}$, $\langle S_{z}
\rangle = -|\langle S_{z} \rangle|$). Therefore, due to the antiferromagnetic
character of the exchange interaction ($J > 0$), $H_{\rm{exch}}$ is also
negative, opposed to $H_{0}$.

The second term in Eq.~\ref{Ham} is the hyperfine interaction between the
(polarized) $\pi$ electrons and Se nuclei
\begin{equation}
\mathcal{H}_{Is} = \sum_{i} \gamma \hbar I_{z}^{i} A_{\pi}^{i}(\theta) g_{\pi}
\mu_{\rm{B}} \langle s_{z}\rangle = \sum_{i} -\gamma \hbar I_{z}^{i}
A_{\pi}^{i}(\theta) \chi_{\pi} H_{\rm{eff}}~,
\end{equation}
where $A_{\pi}^{i}(\theta)$ is the corresponding hyperfine coupling which
depends on the direction in which the field is applied. In our case we will
denote this direction by the angle $\theta$ measuring the rotation of the
sample around the $c$ axis.\cite{zero_angle}

The third term is the dipolar interaction between the localized spins at the Fe
sites and the Se nuclei
\begin{equation}
\mathcal{H}_{IS} = \sum_{i,j} -\gamma \hbar g_{\rm{Fe}} \mu_{\rm{B}}
I_{z}^{i}D_{zz}^{ij}S_{z}^{j} = \sum_{i} -\gamma \hbar I_{z}^{i}
H_{\mathrm{dip}}^{i}(\theta)~,
\end{equation}
where $D_{zz}^{ij}$ is the dipolar coupling tensor and
$H_{\mathrm{dip}}^{i}(\theta)$ the corresponding local dipolar field, which can
be computed exactly since the structure is known.

Putting together all the terms involving nuclear spins, their resonance
frequency $f^{i}$ will be shifted from the reference by
\begin{equation}
\Delta f^{i}\equiv f^{i}-\gamma H_{0}=\gamma [ A_{\pi }(\theta )^{i}\chi _{\pi
}H_{\mathrm{eff}}+H_{\mathrm{dip}}^{i}(\theta )+K_{c}^{i}H_{0} ] ~,
\label{eq:shift}
\end{equation}
where by index $i$ we distinguish 8 Se sites within the unit cell.
As in the low temperature spectra these sites are not resolved, in
the following we consider the average values over $i$ and omit the
index. By subtracting the dipolar and the chemical shift we focus
on the hyperfine contribution of $\pi $ electrons
\begin{equation}
\delta f(H_{0},T)\equiv f-\gamma \lbrack H_{\mathrm{dip}}(\theta,
M_{d}(H_{0},T))+(1+K_{c})H_{0}]=\gamma A_{\pi }(\theta )\chi _{\pi
}[H_{0}-JM_{d}(H_{0},T)]~, \label{hypshift}
\end{equation}
where the exchange field is explicitly written in terms of the Fe moments $%
M_{d}=-g_{\mathrm{Fe}}\mu _{\mathrm{B}}\langle S_{z}\rangle $, whose field
and temperature dependence are given by the (modified) Brillouin function.%
\cite{Wu_2007}  From this equation it is obvious that plotting
$\delta f(H_{0}=$ const.$,T)$\ as a function of $M_{d}(H_{0}=$
const.$,T),$\ with $T$\ being implicit parameter, one obtains a
linear dependence which enables the determination of the $J$\ and
$A_{\pi }(\theta )\chi _{\pi }$\ parameters. This procedure, which
has been used in the previously published work,\cite{Hiraki_2006,
Wu_2007} relies on the predicted temperature dependence of Fe
moments. A more ``robust'' approach is to exploit the field
dependence of this equation in the low temperature limit, where Fe
moments are fully polarized, $M_{d}(H_{0},T\sim 0)=5\mu
_{\mathrm{B}}$, and thus field independent. The same parameters
are obtained from the linear $H_{0}$ dependence of $\delta
f(H_{0},T\sim 0)$.\ Furthermore, by going to high enough field we
can explicitly reach the point where the applied field cancels the
exchange field, that is where the hyperfine shift goes to zero,
$\delta f(H_{0}=5\mu _{\mathrm{B}}J,T\sim 0)=0.$

We remark that it is also easy to study the angular dependence of
hyperfine coupling by comparing the line positions taken at two
well separated field values $H_1$ and $H_2$. As in the low
temperature limit both $H_{\mathrm{dip}}$ and $H_{\mathrm{exch}}$
are field independent, from Eq.~\ref{eq:shift} we get that
\begin{equation}
[\Delta f(H_2) - \Delta f(H_1)]/\gamma (H_2-H_1) -K_{c}= A_{\pi}(\theta ) \chi
_{\pi}~. \label{deltashift}
\end{equation}
Note that the hyperfine coupling $A_{\pi}(\theta )$ is expected to have
uniaxial symmetry reflecting the Se 4$p_{z}$ orbital.

We have seen that the essential prerequisite to discuss the hyperfine shift is
the calculation of the dipolar contribution, including demagnetization effects.
We use the Lorentz method \cite{lorentz} in which we first divide the volume of
the sample into two parts. The first part is a sphere with a radius much larger
than interatomic distance, for which we sum all the individual dipolar fields
assuming a point dipole $M_{d}$ at each Fe site, to obtain the so called
dipolar field $H_{\mathrm{dip}}^{\prime }$. The contribution of the rest of the
sample outside the \textquotedblleft Lorentz sphere\textquotedblright , the so
called demagnetization contribution $H_{\mathrm{dip}}^{\prime \prime }=4\pi
M_{z}(\frac{1}{3}-N)$, is obtained assuming uniform magnetization $M_{z}$
\cite{mz} and the effects of the sample shape are taken into account by the
\textquotedblleft demagnetization factor\textquotedblright\ $N$. \cite{demag}
The total $H_{\mathrm{dip}}$ is the sum of the two terms
$H_{\mathrm{dip}}^{\prime }+H_{\mathrm{dip}}^{\prime \prime }$. Finally, the
chemical shift $K^{\mathrm{c}}$ was taken equal to that measured in
$\lambda$-(BETS)$_{2}$GaCl$_{4}$\cite{takagi2003}, $K^{\mathrm{c}}$ =
-0.0116\%.

\begin{figure}
\begin{center}
\includegraphics[width=11cm]{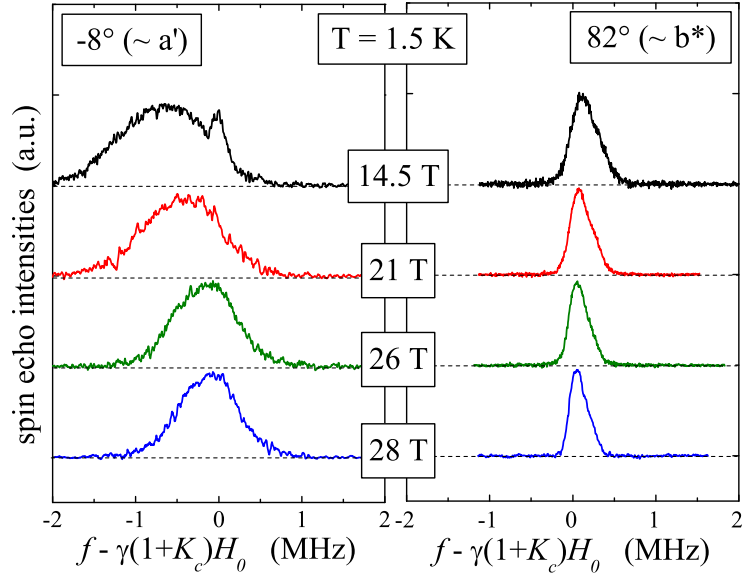}
\end{center}
\caption{$^{77}$Se NMR spectra recorded at four different fixed values of the
applied field $H_{0}$, pointing to two different orientations\cite{zero_angle}
in the $a'b^{*}$ plane. The difference in the shift and linewidth behavior is
due to the anisotropy of the hyperfine coupling $A_{\pi}$ of the conduction
electrons. The decrease of the linewidth on increasing the magnetic field is
due to the decrease of the effective field $H_{\rm{eff}} =  |H_{0} +
H_{\rm{exch}}|$.} \label{spectra_vs_field}
\end{figure}

\section{\label{rd}Results and discussions}


\subsection{\label{j}Low temperature field dependence of the line shift}

Fig.~\ref{spectra_vs_field} shows $^{77}$Se NMR spectra at four values of the
external field from 14.5 to 28 T for two different orientations of the crystal.
For the first orientation (close to $a'$) one clearly observes that on
increasing $H_{0}$ not only the spectrum shifts, but also gets narrower. Such a
decrease of the linewidth is expected if part of it is of magnetic origin and
proportional to $H_{\rm{eff}}$ which decreases as $H_{0}$ increases. This means
that the linewidth is due to a distribution of $A_{\pi}\chi_{\pi}$, pointing to
a possible modulation of $\chi_{\pi}$($\mathbf{R}$). For the second orientation
(close to $b^*$) the variation of shift and the widths of the lines are much
smaller. This difference in the width and shift behavior is actually due to the
anisotropy of the hyperfine coupling $A_{\pi}(\theta)$, as expected for $\pi$
electrons.

\begin{figure}
\begin{center}
\includegraphics[width=11cm]{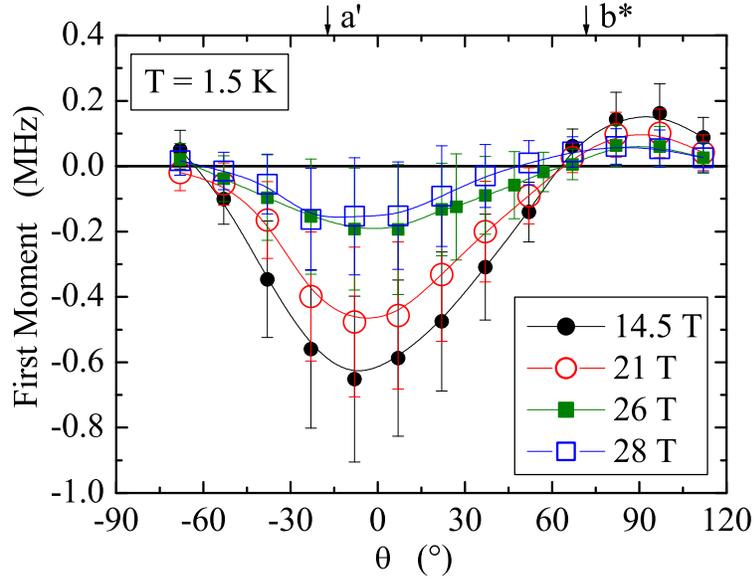}
\end{center}
\caption{Angular dependence\cite{zero_angle} of the first moment of the
$^{77}$Se NMR lines at four different values of $H_{0}$. These data are not
corrected from the dipolar and demagnetization contribution. The vertical bars
accompanying the symbols are not the error bars, but correspond to the width of
the lines. } \label{M1_vs_theta_and_H}
\end{figure}

\begin{figure}
\begin{center}
\includegraphics[width=11cm]{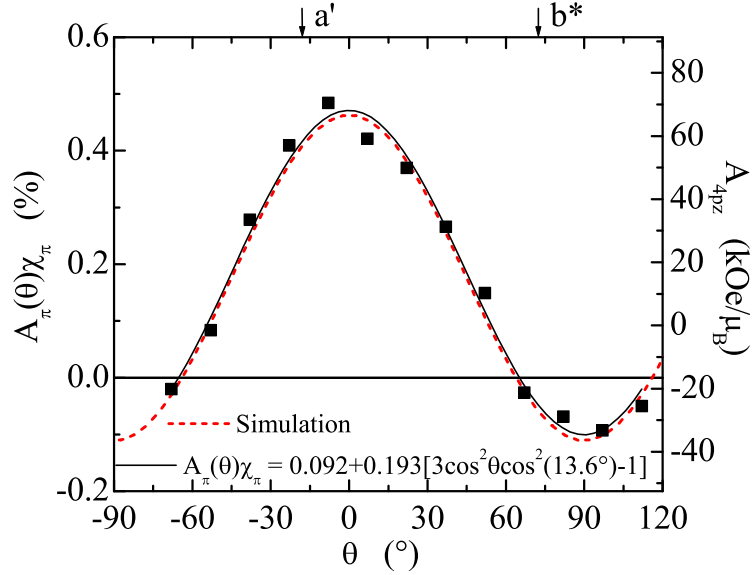}
\end{center}
\caption{ Experimentally determined $A_{\pi}(\theta) \chi_{\pi}$ at 1.5~K
(solid squares), simplified fit for the angular dependence\cite{zero_angle}
(line) and more detailed theoretical prediction (dotted line), as explained in
the text.} \label{A_chi_theta}
\end{figure}

Fig.~\ref{M1_vs_theta_and_H} shows the angular dependence of the shift at  four
field values,  14.5, 21, 26 and 28~T. These raw data include the angular
dependence of dipolar contribution which, because of the demagnetization
factor, is somewhat difficult to obtain exactly, since the shape of the crystal
is not perfectly known. However, it is quite reasonable  to assume that for
these field values the Fe moments are fully saturated at 1.5~K, so that the
dipolar contribution is independent of $H_{0}$. We can thus use
Eq.~\ref{deltashift} to obtain directly $A_{\pi }(\theta)\chi_{\pi}$. From four
available field values, we have created three differential ($H_i - 14.5~T$)
sets of data, and plotted in Fig.~\ref{A_chi_theta} their average $A_{\pi
}(\theta)\chi_{\pi}$ value for each value of $\theta$. The experimental angle
dependence is fitted\cite{zero_angle} to the theoretically expected one,
$\chi_{\pi}[A_{iso} + A_{ax}(3 \cos^{2}\theta \cos^{2}\psi -1)]$, where $\psi$
= 13.6$^{\circ}$ is the (average) minimum angle between the $\pi$ orbitals and
the $a^*b^*$ plane (or $H_0(\theta=0)$ direction\cite{zero_angle}).
From the fit, we found $\chi_{\pi}A_{iso}$ = 0.092~\% and $\chi_{\pi}A_{ax}$ =
0.193~\%. To further determine the value of $A_{\pi}$ we need an estimate for
$\chi_{\pi}$. We will assume that $\chi_{\pi}$ in $\lambda$(BETS)$_2$FeCl$_4$
is the same as in $\lambda$(BETS)$_2$GaCl$_4$ where $\chi_{\pi}$~=
4.5$\times$10$^{-4}$~emu/mole at room temperature and
6.3$\times$10$^{-4}$~emu/mole at low temperature.\cite{tanaka1999} Retaining
the low temperature value of $\chi_{\pi}$, we find $A_{ax}=$
35~kOe/$\mu_{\rm{B}}$. This value should be compared to a theoretical
prediction for the dipolar hyperfine coupling constant of a $p$-type (4$p_z$)
electronic orbitals, $A_{ax} = \frac{2}{5}\langle r^{-3}
\rangle\mu_{\mathrm{B}}\sigma$, where for $\pi$ orbitals of Se in the BETS
molecule\cite{Fraga} $\langle r^{-3}\rangle =$ 9.28~$a_{0}^{-3}$ and the spin
density\cite{takagi2003} is $\sigma \sim$~0.16. We find the predicted value
$A_{ax}$~= 37~kOe/$\mu_{\rm{B}}$ in excellent agreement with the experimental
value of 35~kOe/$\mu_{\rm{B}}$ estimated above. In the analysis given above we
assumed that all the $\pi$ orbitals point in the same direction in space. As
this is only approximately true, we also calculated the predicted angular
dependence of the hyperfine coupling considering the exact orientation of
orbitals for all 8 Se sites in the molecule. The average hyperfine coupling
($\langle A_{\pi}^i(\theta) \rangle_i$) obtained in this way, plotted in
Fig.~\ref{A_chi_theta}, is nearly indistinguishable from the simplified fit,
confirming excellent agreement with the theory.

\begin{figure}
\begin{center}
\includegraphics[width=11cm]{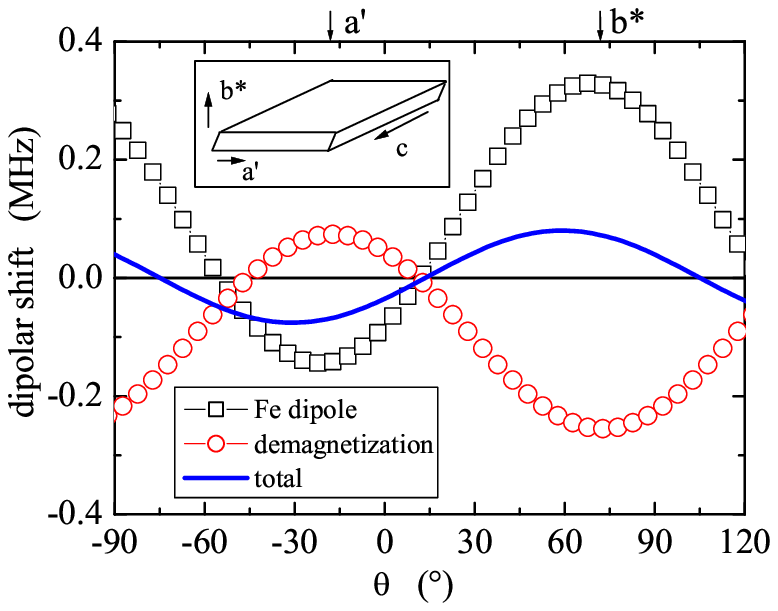}
\end{center}
\caption{ Contribution of the dipolar interaction and of the demagnetization
field (N$_{a'}$ = 0.2, N$_{b*}$ = 0.8, N$_{c}$ = 0) as a function of the angle
in the $a'b^{*}$ plane. Inset: shape of the crystal.} \label{demag}
\vspace{1.0cm}
\begin{center}
\includegraphics[width=11cm]{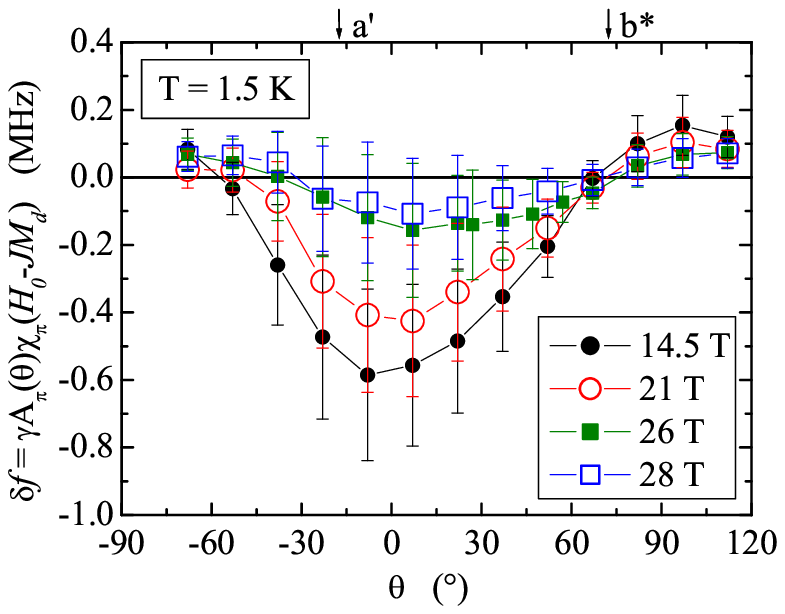}
\end{center}
\caption{Angular dependence of the shift of the Se lines at 14.5, 21, 26 and
28~T after substraction of the dipolar and demagnetization contribution.}
\label{theta2}
\end{figure}

We shall now come to the problem of the demagnetization factor. In
Fig.~\ref{demag} is shown the predicted angular dependence of the dipolar
contribution to the lineshift from a Lorentz sphere of 35~\AA~radius, the
demagnetization contribution due to the shape of the crystal, and their sum. As
the sum has quite large amplitude, it is expected to play an essential role in
the determination of $H_{\rm{exch}}$. We also remark that the sum results from
a strong compensation of two components which are both very strong. As the
shape of the tiny crystal and thus the demagnetization factor is not precisely
known, this can introduce some uncertainty in the determination of
$H_{\rm{exch}}$.

\begin{figure}
\begin{center}
\includegraphics[width=11cm]{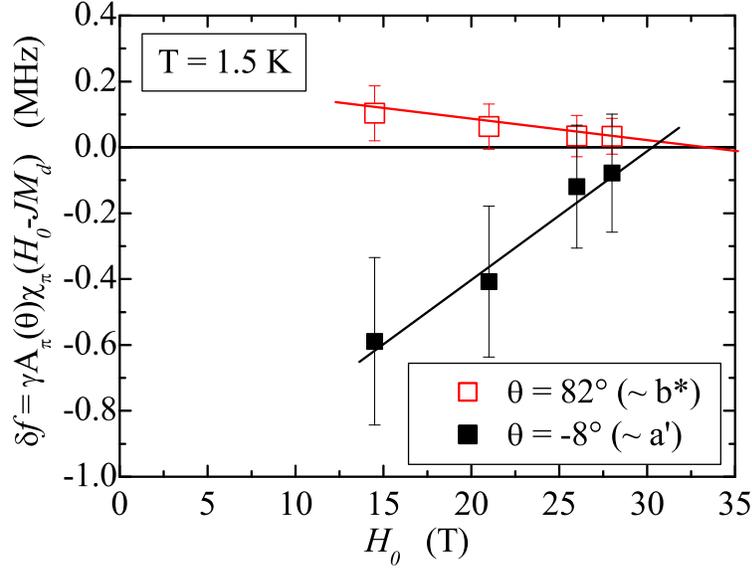}
\end{center}
\caption{Determination of the exchange field from the field variation of the
shift at two different values of $\theta$. The $|H_{\rm{exch}}|$ is given by
the intercepts of the shift with the zero value and found equal to
~32~$\pm$~2~T.} \label{H_exch_field}
\end{figure}

In Fig.~\ref{theta2}, is shown the angular variation of the shift of the Se
lines after correction from the dipolar and the demagnetization contributions,
for the four different magnetic field values. This corrected experimental
values correspond to $\delta f$ defined by Eq.~\ref{hypshift}. As explained in
Section~\ref{nmr} below Eq.~\ref{hypshift}, a linear fit of $\delta f$ as a
function of $H_{0}$ shown in Fig.~\ref{H_exch_field} allows direct
determination of $H_{\rm{exch}}$ since $\delta f = 0$ corresponds to $H_{0} +
H_{\rm{exch}} = 0$. We obtain $|H_{\rm{exch}}|$~=~32~$\pm$~2~T, in very good
agreement with the value of 33~T corresponding to the maximum of $T_{c}$. We
underline that this value is obtained without any assumption on the values of
the hyperfine field $A_{\pi}(\theta)$ and the susceptibility $\chi_{\pi}$ of
the $\pi$ band. The main source of error is the determination of
demagnetization contribution. That a small error of that type is possible can
be seen in Fig.~\ref{theta2}: although the extrema of the shift should appear
at the same value of $\theta$ whatever is the applied magnetic field, one can
observe a slight deviation, which could indicate that our correction for the
dipolar contribution is not completely correct.

\subsection{\label{k}Temperature dependence of the line shift}

\begin{figure}
\begin{center}
\includegraphics[width=11cm]{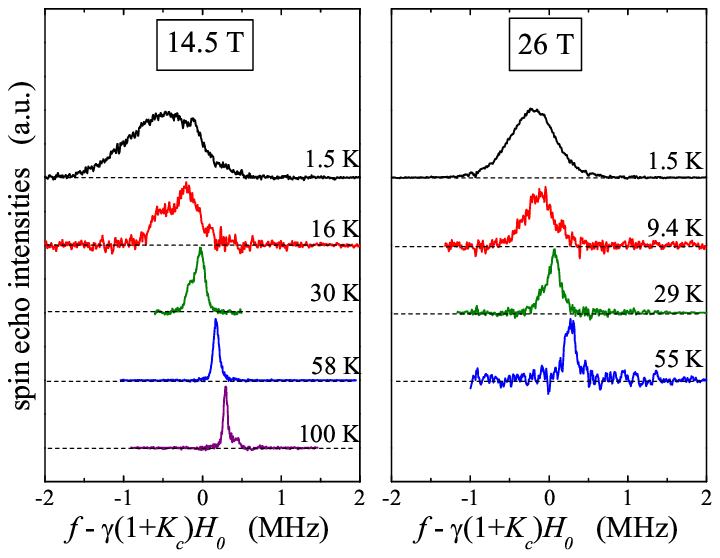}
\end{center}
\caption{ Se NMR spectra at various temperatures at 14.5 and 26 T. The field
direction is close to  $a'$.} \label{spectra_temp}
\vspace{1.0cm}
\begin{center}
\includegraphics[width=11cm]{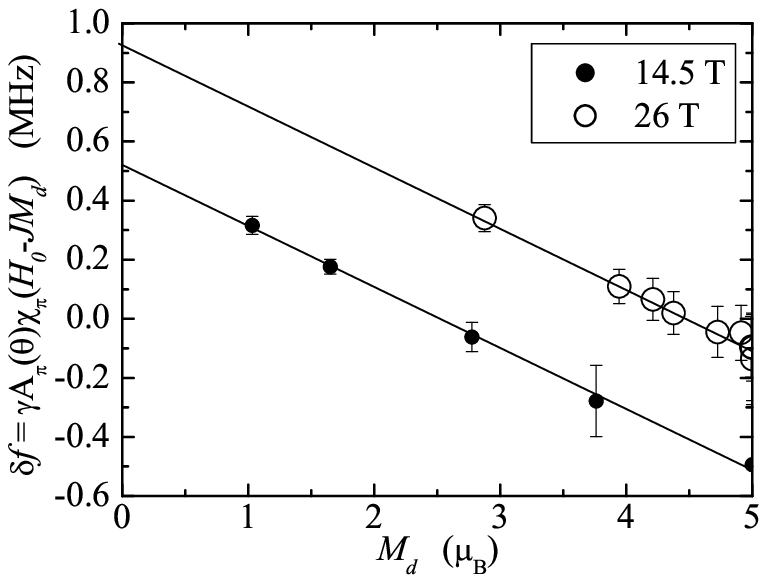}
\end{center}
\caption{ Resonance shift as a function of the Fe moment $M_{d}$ which is
calculated by a modified Brillouin function taken from \cite{Wu_2007}.}
\label{brillouin}
\end{figure}

As we have seen in section \ref{nmr}, the value of $H_{\rm{exch}}$ can also be
extracted from the temperature dependence of the shift of the Se line at
constant value of the field, provided one knows the temperature dependence of
$M_{d}$. In a previous paper, \cite{Hiraki_2006} we have analyzed in this way
our results obtained at 14.5~T. However, for the determination of
$H_{\rm{exch}}$, we did not take into account the demagnetization factor, and
we considered the Brillouin function of independent Fe moments. As shown by Wu
\emph{et al.},\cite{Wu_2007} this Brillouin function has to be modified to take
into account the effect of antiferromagnetic interactions between these Fe
moments, through the $\pi$ conduction electrons. Here we present such complete
analysis for two values of the field, $H_0$~= 14.5 and 26~T. In
Fig.~\ref{spectra_temp} are shown the temperature dependence of the Se NMR
lines with temperature at these two fields. Evaluating the temperature
dependence of $M_{d}$ using the modified Brillouin function mentioned above,
and correcting the data for the demagnetization field and dipolar field, one
can plot the corrected shift which is proportional to $\gamma A_{\pi}(\theta%
)\chi_{\pi}[H_{0}-JM_{d}]$ as a function of $M_{d}(T)$ with the temperature as
an implicit parameter, as shown in Fig.~\ref{brillouin}. The intersection of
the straight lines on the vertical axis ($M_{d}~=~0$) corresponds to $\gamma
A_{\pi}(\theta)\chi_{\pi}H_{0}$, leading to the estimate
$A_{\pi}(\theta)\chi_{\pi}$~= 0.44~\% close to the value of 0.37~\% determined
in the previous subsection (for the corresponding orientation) by a method
which does not depend on any evaluation of demagnetization factor nor on the
temperature dependence of $M_{d}$. From the slope of the lines in
Fig.~\ref{brillouin} and the above given two $A_{\pi}(\theta)\chi_{\pi}$
values, we obtain $|H_{\rm{exch}}|$~= 30 and 34~T. The average of these two
values, $|H_{\rm{exch}}| = 32 \pm 2$~T, is the same as the estimate obtained in
the previous subsection, in agreement with the expected value of 33~T
corresponding to the maximum of the transition temperature for
superconductivity.

\subsection{\label{cd}Anomalous broadening of NMR line}

\begin{figure}
\begin{center}
\includegraphics[width=11cm]{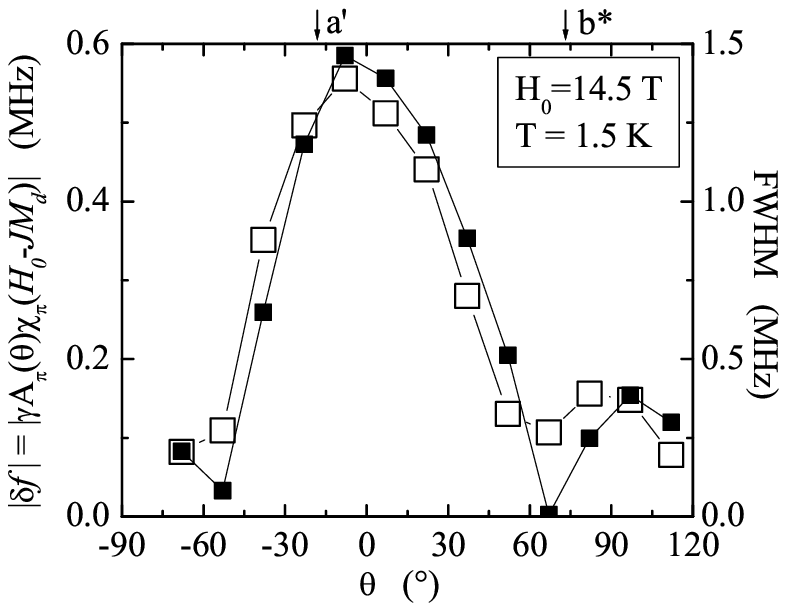}
\end{center}
\caption{ Absolute value of the shift corrected for the dipolar contribution
($|\delta f|$, closed symbols, left scale) and linewidth (open symbols, right
scale) of the $^{77}$Se NMR line as a function of the field orientation in the
$a'b^{*}$ plane at 14.5~T and 1.5~K. Note that the angle at which the linewidth
is minimum ($\theta \simeq 64^{\circ}$) is also the one at which
$A_{\pi}(\theta)$ vanishes as determined from Fig.~\ref{A_chi_theta}.}
\label{shiftwidth}
\end{figure}

As shown in Fig.~\ref{spectra_temp}, very broad NMR lines were observed for
$H_{0} \parallel$ \textit{a'} at low temperature. This anomalous line
broadening was observed only in the low temperature region ($T<$ 30~K), while
at higher temperatures the linewidth scales to the $M_{d}$. To find the
mechanism of this low temperature line broadening in the metallic state, we
have measured the angular dependence of the NMR spectrum when $H_{0}$ is
rotated in the $a'b^{*}$ plane at 1.5 K. Fig.~\ref{shiftwidth} presents both
the linewidth (full width at half maximum, FWHM) and the shift data. It turns
out that the angular dependence of the width is strongly correlated to that of
the shift. The maximum of the width and the (negative) extremum of the shift
nearly coincide at $\theta \simeq 0$. Also the minimum of the width is observed
for the direction which gives a zero shift, that is $\theta \simeq 64^{\circ}$.
This strongly suggests that the line broadening is not due to defects in the
crystals, but is caused by an intrinsic spatial distribution of the spin
susceptibility $\chi_{\pi}(\mathbf{R})$. This is also supported by the field
dependence of the linewidth shown in Fig.~\ref{spectra_vs_field}, as already
mentioned in the beginning of Section~\ref{j}. Higher $H_0$ values correspond
to a smaller value of $H_{\mathrm{eff}}$ and thus to smaller linewidth. Let us
call $\Delta \chi_{\pi}$ the second moment of spin susceptibility distribution.
From Fig.~\ref{shiftwidth}, $\Delta \chi_{\pi}$ is comparable to $\chi_{\pi}$.
As the spectra are just broadened without any appreciable structure, this
indicates that the $\Delta \chi_{\pi}$ is continuously distributed in the
crystal. Supposing that the spin density  is proportional to the charge density
in the paramagnetic state, one possible mechanism for this anomalous broadening
is charge disproportionation in the conducting layer which has been proposed to
explain  microwave conductivity\cite{matsui2001} and X-ray
measurements.\cite{komiyama2004} This would indicate that the distribution of
the BETS valence is remarkably large.
Quite similar line broadening has been observed in the charge ordering system, $\theta$-(BEDT-TTF)$_{2}$RbZn(SCN)$_{4}$ in the ``metallic'' state above the metal-insulator transition temperature.\cite{chiba2004}
Another possible mechanism for the broadening is an
oscillation of spin polarization in the conducting layer induced by the
magnetic moment of the Fe atoms. This problem should be addressed in more
details in the future.

\section{Concluding remarks}

We have performed $^{77}$Se NMR measurements in the Field Induced
Superconductor, \mbox{$\lambda$-(BETS)$_{2}$FeCl$_{4}$} in the low temperature
and high magnetic field metallic regime. Our study allowed a direct microscopic
determination of the exchange field $H_{\textrm{exch}}$ induced by the exchange
coupling between the conduction electrons of the $\pi$ band and the localized
spin S = 5/2 of the Fe atoms. Two independent ways have been used. One is based
on the temperature dependence of the Fe moment and that of the shift of the
$^{77}$Se NMR line. Another method was to measure at low temperature, so that
the Fe moments are saturated, and to record the angular dependence of the shift
at (four) different applied field values varying from 14.5 to 28~T. This
allowed us to determine directly the product of the susceptibility of the $\pi$
band and the hyperfine coupling, $A_{\pi}(\theta)$, which contains an isotropic
term and one corresponding to the spin polarization of $\pi$ orbitals at the Se
sites. As the external field increases, the amplitude of the angular variation
of the shift decreases, since it is proportional to the total effective field
$H_{0} + H_{\textrm{exch}}$ in which two components have opposite signs. This
is a direct experimental proof of the compensation mechanism proposed by
Jaccarino and Peter.\cite{jaccarino1962} Both methods led to
$|H_{\textrm{exch}}| = 32 \pm 2$~T, in excellent agreement with the value of
33~T for which the transition temperature of the superconducting phase is
maximum. This value is also in agreement with theoretical
estimates.\cite{cepas2002,mori2002} The error bars are mainly due to the
difficulty to evaluate the demagnetization field in our sample.

In addition to these main findings, an anomalous line broadening has been
observed at low temperatures. The linewidth has been found proportional to the
hyperfine shift and to have the same angular dependence. One possible
interpretation is the occurrence of charge disproportionation in the metallic
state neighboring the superconducting phase. The relationship between these two
types of order has  been recently discussed in the charge ordering system
$\alpha$-(BEDT-TTF)$_2$I$_3$.\cite{tajima2002,kobayashi2005}.

\section{Acknowledgment}
The authors are grateful to Prof. K. Takimiya from Hiroshima University for the
$^{77}$Se isotope enrichment of the sample and to Prof. W.G.~Clark from
University of California, Los Angeles, for advises about tiny NMR coil. This
work was partially supported by Japan Society for Promotion of Science, by High
Technical Research Center of Gakushuin University, by Grant-in-Aid for
Scientific Research on Priority Areas of Molecular Conductors (No. 15073221)
from the Japanese Ministry of Education, Culture, Sports, Science and Technology,
and by the French ANR grant 06-BLAN-0111.


\begin{thebibliography}{9}
\bibitem{zero_angle} The zero reference for the angle $\theta$ has been taken
to correspond to the maximum of the fit given in Fig.~\ref{A_chi_theta}.
Physically this corresponds to magnetic field being closest to the (average)
direction of electronic $\pi$ orbitals. With respect to this reference $a'$
direction (lying in the $ac$ plane) is at $\theta(a') = -17^{\circ}$.
\bibitem{demag}
Considering the dimensions of the sample, we found N$_{a'}$ = 0.2,
N$_{b*}$ = 0.8, N$_{c}$ = 0.
\end{thebibliography}
\end{document}